\documentclass[letterpaper, 10 pt, conference]{ieeeconf}  % Comment this line out if you need a4paper

\IEEEoverridecommandlockouts                              % This command is only needed if 
\usepackage{graphics} % for pdf, bitmapped graphics files
\usepackage{subcaption}
\usepackage{epsfig} % for postscript graphics files
\usepackage{amsmath} % assumes amsmath package installed
\usepackage{amssymb}  % assumes amsmath package
\usepackage[short]{optidef}
\usepackage{ar}
\usepackage{bm}
\usepackage{hyperref}
\usepackage{ar}
\usepackage{color}

\usepackage[utf8]{inputenc}
\usepackage{tikz}
\usepackage{pgfplots}
\newlength\figureheight
\newlength\figurewidth
\usepgfplotslibrary{groupplots}

\usepackage{pythontex}
\begin{pycode}
import sys
import os
def add_to_path(p):
	if p and p not in sys.path:
		sys.path.append(p)
\end{pycode}
\setpythontexworkingdir{.}
\setpythontexoutputdir{.}
\setpythontexcontext{textwidth=\the\textwidth, textheight=\the\textheight}

\newif\ifcomment
\setboolean{comment}{false}

\ifthenelse{\boolean{comment}}{\newcommand{\jochem}[1]{\textcolor{red}{(Jochem: #1)}}}{\newcommand{\jochem}[1]{}}

\newcommand{\barr}{\begin{array}}
\newcommand{\earr}{\end{array}}

\usepgfplotslibrary{external} 
\title{\LARGE \bf Vertical Airborne Wind Energy Farms with High Power Density per Ground Area based on Multi-Aircraft Systems}

\author{Jochem De Schutter$^{1}$, Jakob Harzer$^1$, Moritz Diehl$^{1,2}$% <-this % stops a space
\thanks{$^{1}$Systems Control  and Optimization Laboratory, Department of Microsystems Engineering (IMTEK) and Department of Mathematics, University of Freiburg, Georges-Koehler-Allee 102, 79110 Freiburg, Germany.
	{\tt\small jochem.de.schutter@imtek.de, jakob.@imtek.uni-freiburg.de, moritz.diehl@imtek.uni-freiburg.de}}%
\thanks{$^{2}$Department of Mathematics, University of Freiburg, Ernst-Zermelo-Strasse 1, 79104 Freiburg im Breisgau, Germany.}
}

\usepackage{booktabs}
\begin{document}

\maketitle

\begin{abstract}
This paper proposes and simulates vertical airborne wind energy (AWE) farms based on multi-aircraft systems with high power density (PD) per ground area.
These farms consist of many independently ground located systems that are flying at the same inclination angle, but with different tether lengths, such that all aircraft fly in a large planar elliptical area that is vertical to the tethers.
The individual systems are assigned non-overlapping flight cylinders depending on the wind direction.
Detailed calculations that take into account Betz' limit, assuming a cubically averaged wind power density of 7 m/s, give a potential yearly average PD of 43 MW/km$^2$.
A conventional wind farm with typical packing density would yield a PD of 2.4 MW/km$^2$ in the same wind field.
More refined simulations using optimal control result in a more modest PD of 6 MW/km$^2$ for practically recommended flight trajectories.
This PD can already be achieved with small-scale aircraft with a wing span of 5.5 m.
The simulations additionally show that the achievable PD is more than an order of magnitude higher than for a single-aircraft AWE system with the same wing span.

\end{abstract}
\section{Introduction}
Because of the abundant availability of wind and solar energy resources, in principle only a tiny fraction of the earth's surface area would suffice to generate all of humanity's energy needs.
Nevertheless the power density (PD) per ground surface area of wind and solar power technologies is still a relevant quantity, since the infrastructure costs of renewable energy farms, such as grid connection and installation logistics, scale proportionally with the farm area.
The power density of
existing wind power farms is estimated to be around
PD = 2 MW/km$^2$. 
For solar PV farms we have a PD of around 10 MW/km$^2$ \cite{VanZalk2018}.

Airborne wind energy (AWE) is an upcoming renewable energy technology which aims at harvesting the steady and strong high-altitude winds that cannot be reached by conventional wind technology, at only a fraction of the resources.
AWE developers mainly consider single-aircraft AWE systems (S-AWES), which are based on the principle of one tethered aircraft flying fast crosswind maneuvres.
However, S-AWES are subject to several limitations that impede the technology to increase PD with respect to conventional wind.

First, S-AWES are characterized by high tether drag dissipation losses.
These losses are inversely proportional to the aircraft size, which is why large and heavy (and thus, costly) aircraft are needed to achieve the efficiency needed for a high PD. 
Second, large aircraft come with an inherently large turning radius, which corresponds to a large trajectory footprint on the ground.
Therefore, a very dense geometric spacing of units - possibly with shared airspace - needs to be achieved for high PD.
Third, since the maximum tether length is limited due to the drag losses, S-AWES in park configuration would all operate at similar altitudes, which, in combination with close geometric spacing, would lead to large losses due to wake interaction \cite{Haas2022}.

\begin{figure}
	\centering
	\input{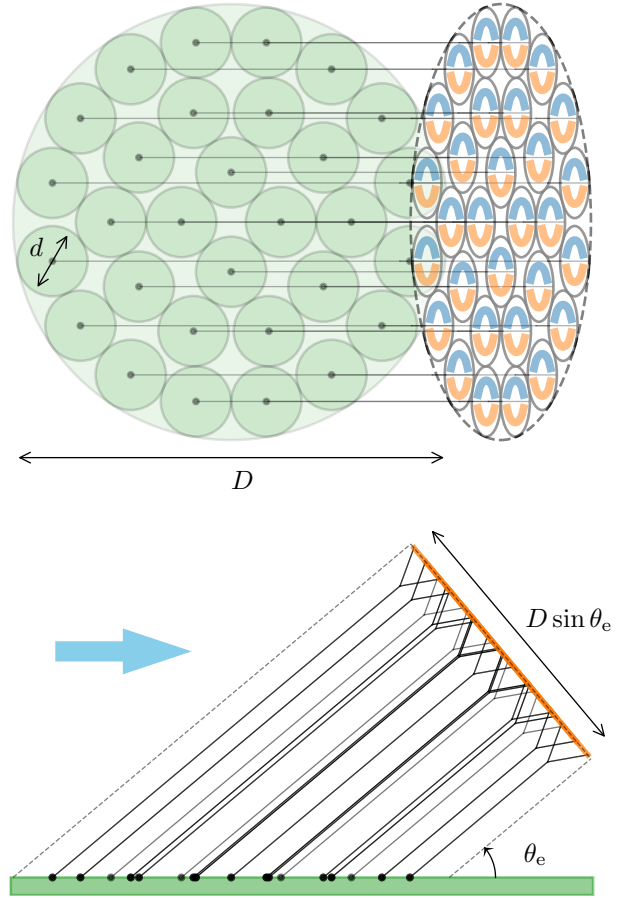}
	\caption{Top and side view of a vertical M-AWES wind farm with $N = 30$, $\theta_{\mathrm{e}} = 40^{\circ}$, $d/D = 0.16$ and $l_{\min} / D = 0.42$.}\label{fig:example_farm}
\end{figure}

To overcome these limitations, this work proposes (and simulates) the concept of vertical AWE farms based on multi-aircraft AWE systems (M-AWES), as depicted in Fig~\ref{fig:example_farm}.
In M-AWES, two or more aircraft fly very tight loops around a shared, quasi-stationary main tether \cite{Zanon2013a}.
Therefore, M-AWES are very efficient even for small aircraft size while using airspace more effectively resulting in a lower trajectory footprint on the ground.
The proposed vertical M-AWES farm layout additionally exploits the fact that M-AWES can fly at arbitrarily high locations above the ground, so that they can operate at distinct locations in the sky, thereby avoiding wake interaction.
A somewhat similar idea based on networked rotary AWE systems was proposed (but not simulated) in \cite{Read2018}.

The remainder of this text is structured as follows.
Section \ref{sec:main_idea} describes the vertical M-AWES park concept and determines an upper PD limit based on a simplified analysis.
Section \ref{sec:system_model} states the system model used in the simulation study while Section \ref{sec:OCP} proposes an optimal control problem (OCP) formulation to compute PD-optimal power cycles.
Section \ref{sec:results} describes the numerical results of a case study, where we compute PD-optimal orbits for both S- and M-AWES, for a small and moderate aircraft size, and where we investigate the trade-off between PD and wing area efficiency.
Section \ref{sec:discussion} discusses the main conclusions and gives suggestions for future research.

%Assuming very close geometric spacing, estimates for this technology range from up to 4 MW/km$^2$ for moderate-size systems to up to 19.6 MW/km$^2$ for large-size systems \cite{Weber2021}.
%These estimates however do not take into account wake interaction, which is significant for dense AWE parks with large-size systems \cite{Haas2022}.

%This work proposes and simulates circular AWE farms
%with high PD. 
%Detailed calculations that take into account Betz’ limit
%show that the effective wind harvesting area is propor-
%tional to the ground area of the farm for this concept, but
%reduced by several factors: first the circle packing loss fac-
%tor ηcirc = 0.7, second the cosine and area reduction loss
%factor ηgeo = cos(α)2 sin(α) = 0.375, third the Betz factor
%ηBetz = 16/27. Together, they lead to an effective area re-
%duction factor ηtotal = ηcirc ηgeo ηBetz = 0.15. Assuming
%a cubically averaged wind speed of v = 7 m/s, the yearly
%average wind power density in the air would be given by
%pair = 1/2ρairv3 = 206 W/m2, such that the PD of the AWE
%farm would be given by pPD = ηtotalpair = 32 MW/km2,
%which is more than 15 times the PD for conventional wind
%and 3 times the PD of solar PV. This assessment is sup-
%ported by more detailed simulation studies of a wind farm
%consisting of many moderately sized dual-kite systems.

\section{Vertical M-AWES parks}\label{sec:main_idea}

The vertical M-AWES parks proposed in this work are circular and consist of many independently ground located M-AWES that are all
flying at the same tether inclination angle $\theta_{\mathrm{e}}$, but
with different tether lengths, depending on the wind direction, such that all systems fly in a large planar elliptical area that is vertical to the tethers, as shown in Fig~\ref{fig:example_farm}.

The individual systems are assigned non-overlapping ``operation cylinders" that depend on the wind direction and correspond to a circular ground area.
Otherwise, the individual systems are completely independent and can e.g. be started and landed independently.
Each system might be located on a small tower of e.g. 5 m height in order to minimize its impact on the usability of the ground area for agriculture.

%Idea is a circular shaped park with a diameter of $D=2$ km, i.e. a ground area of $3.141$ km$^2$ consisting of many independent dual-kite systems.  
The distance from one system to all others on the ground is lower bounded by the maximum diameter $d$ of each system’s ground area circle.
The optimal circle packing density of $N$ circles in a large circle of diameter $D$ depends on $N$ and is assumed to be $\rho_{\mathrm{circle}}=70\%$ here.
Thus, from now on we assume that
\begin{equation}
N \frac{\pi d^2}{4} = \rho_{\mathrm{circle}} \frac{\pi D^2}{4}
\end{equation}
which means that we choose to build altogether $N= \rho_{\mathrm{circle}} D^2/d^2$ individual units in the park. 
This abstracts from the individual system’s size $d$ and number $N$, such that we only need to remember the packing loss factor  $\rho_{\mathrm{circle}}$.

%Each system operates at the same main tether inclination angle $\alpha=30^\circ$, but the systems have varying tether lengths, depending on their location in the park with respect to the wind direction. The longest tethers (2km) are needed for the upwind locations, the shortest (337 m) for the downwind locations. 
Together, the systems form a large inclined elliptical area in the sky, and all wings fly in this elliptical area.
The area is perpendicular to the tethers, and thus forms an ellipse with a maximum width that is equal to $D$, but a minimum width of $D \sin \theta_{\mathrm{e}}$.
The M-AWES are assumed to be able to fly ellipses with approximately this aspect ratio so the circle packing from the ground can be mapped by an affine transformation to the ellipse packing in the sky. 
The elliptical area forms an angle of $\theta_{\mathrm{e}}$ with the vertical, and thus, the effective area of this ``actuation ellipse” is again reduced by a factor $\cos \theta_{\mathrm{e}}$ resulting in a height of the inclined ellipse of only $D \cos \theta_{\mathrm{e}} \sin\theta_{\mathrm{e}}$.

The shortest tether length $l_{\min}$ defines the location of the lowest point of the ellipse, which is located $l_{\min}\sin \theta_{\mathrm{e}}$ above the ground and $l_{\min} \cos \theta_{\mathrm{e}}$ downwind, extending the ground boundaries of the wind park. 
This causes an extended park diameter $D + 2 l_{\min} \cos \theta_{\mathrm{e}}$, defining a circle above which the AWE systems can fly. 
This virtual area enlargement does not lead to increased infrastructure costs and is therefore neglected.
% Note that the height of obstacles in the ring around the wind park is limited by $\sin \alpha$ times the distance from the wind park, i.e., at a distance of 50m from the park one could accept already 25m tall trees. Note that the numbers are chosen such that the highest AWE system has a height of exactly $1km$ above the ground station locations.

%do not operate at Loyd’s limit, but that they fly slower, e.g. at a wing-wind-speed ratio of $\lambda=7$, which is typical for large wind turbines and leads to accepable noise levels. 
We assume that the M-AWES wing area is adapted to reach the Betz limit $\eta=16/27$ on the available flight area assigned to each individual system.
%The so called cosine-losses for AWE systems (cf. Diehl2013a) are indirectly accounted for in the geometrical area reduction considerations that follow and are therefore not explicitly included.
%The wind power that each system can harvest is proportional to its ground occupation, i.e., the circle area  $\pi d^2 /4$, and the wind power density $p=1/2 \rho_{\mathrm{air}} v^3$. 
The overall power of the wind park is then proportional to its ground area, but affected by a variety of losses: 
\begin{itemize}
	\item the circle packing loss $\rho_{\mathrm{circle}}=70\%$;
	\item the geometric area reduction efficiency that reaches a maximum value of $\cos \bar{\theta}_{\mathrm{e}} \sin\bar{\theta}_{\mathrm{e}} = 0.5$ for $\bar{\theta}_{\mathrm{e}} = 45^{\circ}$;
	\item the Betz factor $\eta=16/27$;
\end{itemize}
leading altogether to an effective loss of
\begin{equation}
 \eta_{\mathrm{tot}} =  \eta\cos \bar{\theta}_{\mathrm{e}} \sin\bar{\theta}_{\mathrm{e}}\rho_{\mathrm{circle}} = 0.3 \rho_{\mathrm{circle}} = 21\% \ .
\end{equation}
The maximum power density is then given by
\begin{equation}
\mathrm{PD}_{\mathrm{max}} = \frac{1}{2}\eta_{\mathrm{tot}} \rho_{\mathrm{air}}v^3.
\end{equation} 
For example, for a wind speed of $v = 7 \ \mathrm{m/s}$ and $\rho_{\mathrm{air}} = 1.2 \ \mathrm{kg/m}^3$ this results in a $\mathrm{PD}_{\mathrm{max}}$ of 43 MW/km$^2$.  

To obtain an estimate for a conventional wind energy farm operating in the same wind field, we assume operation at Betz' limit and a circular packing with a distance of at least 6 rotor diameters between the systems.
There are no geometric area reduction losses, resulting in a total efficiency $\eta_{\mathrm{tot}}=1.2 \%$ and a $\mathrm{PD}_{\mathrm{max}}$ of 2.4 MW/km$^2$ when wake losses are ignored.
This PD is a factor 17 lower than the potential PD of the vertical M-AWES farm.

%Thus, the available power per ground area of the park would be given by $\eta_{\mathrm{total}}  p =37$W/m$^2$ or equivalently by a GAPD of 
%The park with a diameter of $D=2$km has an area of 3.141 km$^2$ and thus generates an average power of 116 MW. Given the typical range of wind speed variations, the nominal installed capacity of the park might be about five times this value, say 500 MW, and the park’s power output will thus vary between 0 MW and 500 MW. 

\section{System model}\label{sec:system_model}

To make a more realistic assessment, we will compute detailed, PD-optimal orbits for individual S- and M-AWES in this work.
We consider ``lift-mode" AWE systems, where power is produced in a periodic fashion: first, the tether is reeled out at high tension, driving a winch at the ground station.
Then, tether is reeled back in again at low tension, resulting in a net positive energy gain.
This section presents the M-AWES dynamics used in the optimal control problem formulation in Section \ref{sec:OCP} and introduces an averaged induction model to account for the Betz losses.

\subsection{Multi-aircraft dynamics}
In the following simulations, we use the multi-aircraft model structure and model parameters described in \cite{DeSchutter2019}.
The system dynamics model all six degrees-of-freedom of the aircraft in the system.
The tethers are assumed to be straight and inelastic, which is a good assumption when tether tension is high.
The dynamics are expressed in non-minimal coordinates and summarized by the implicit DAE
\begin{equation}
F(\dot{x}(t), x(t), u(t), z(t), \theta, a) = 0
\end{equation}
and consistency conditions $C(x(t)) =  0$.

The system variables consist first of the state vector $x \in \mathbb{R}^{n_{\mathrm{x}}}$.
The control vector $u \in \mathbb{R}^{n_{\mathrm{u}}}$ consists of the aircraft aileron, elevator and rudder deflection rates as well as the tether reeling acceleration.
The algebraic state $z \in \mathbb{R}^{n_{\mathrm{z}}}$ consists of the Lagrange multipliers related to the constraints that define the interlinkage of aircraft and tethers.
The system parameters $\theta \in \mathbb{R}^{n_{\theta}}$ represent parameters that can be optimized over, such as the main tether diameter and, in the M-AWES case, the secondary tether length and diameter.
The variable $a \in \mathbb{R}$ is the average induction factor that will be explained in Section \ref{sec:induction}.

For the sake of brevity, we refer the reader to \cite{DeSchutter2019} for a complete and formal description of the system variables, dynamics, aerodynamic forces, consistency conditions, etc.
Here we only explicitly discuss those model components relevant for the wind power availability.
We model the wind shear in a simplified way with a power law approximation:
\begin{equation}
u_{\infty}(z) = u_{\mathrm{ref}}\left( \frac{z}{z_{\mathrm{ref}}}\right)^{c_{\mathrm{f}}} \ ,
\end{equation}
with $u_{\infty}(z)$ the freestream velocity at altitude $z$ and $u_{\mathrm{ref}}$ the reference wind speed measured at altitude $z_{\mathrm{ref}} = 100$ m, with $c_{\mathrm{f}} = 0.15$ a surface friction coefficient typical for flat, open terrain.

The atmospheric density drop with altitude is modeled using the international standard atmosphere model \cite{Archer2013a}:
\begin{equation}
\rho(z) \coloneqq \rho_{\mathrm{0}} \left(\frac{ T_0 - T_{\mathrm{L}}z}{T_0}\right)^{\frac{g}{T_{\mathrm{L}}R}-1}\ ,
\end{equation}
where $R$ is the universal gas constant.
The parameters $T_0$ and $\rho_0$ are the temperature and air density at sea level, and $T_{\mathrm{L}}$ is the temperature lapse rate.

The model is based on a validated, small aircraft model with wing span $b_{\mathrm{ref}} = 5.5 \ \mathrm{m}$, mass $m_{\mathrm{ref}} = 36.5 \ \mathrm{kg}$ and inertia tensor $J_{\mathrm{ref}}$ given in \cite{Malz2019}.
In order to be able to evaluate the dynamics also for larger wing spans $b$, we utilize the following mass upscaling formula:
\begin{align}
m & = m_{\mathrm{ref}} \left(\frac{b}{b_{\mathrm{ref}}}  \right)^{\kappa} \ , \quad \text{and} \quad J = J_{\mathrm{ref}} \left(\frac{b}{b_{\mathrm{ref}}}  \right)^{\kappa+2} \ ,
\end{align}
with upscaling exponent $\kappa = 2.4$.
\subsection{Induction model} \label{sec:induction}
State-of-the-art induction models for AWE are typically a variation of Betz' analysis for conventional wind turbines, thus based on a steady-state analysis \cite{DeLellis2018a, Kheiri2019}.
However, detailed  wind field simulations \cite{Haas2022} show that for lift-mode AWE systems, induction is inherently time-dependent: at the beginning of the reel-out phase, induction starts to build up, reaching its peak when transitioning into the reel-in phase, after which it starts to decline.
When the power cycle re-starts, the wind is almost ``fresh" again.

While the induction model proposed in \cite{Leuthold2018} accounts for the dynamic variability of the aircraft trajectories over a power cycle, it still assumes an instantaneous build-up of induction.
Therefore, in this work, we propose the following model, based on a momentum balance applied to averaged flight quantities.

We first compute the annular swept area for each aircraft $k \in \mathcal{K}$, with $\mathcal{K}$ the index set of all aircraft in the system.
We integrate over the reel-out phase the norm of the aircraft's flight speed $\dot{q}$ multiplied with the wing span, weighted with the local dynamic pressure:
\begin{equation}
A_{\mathrm{s},k} \coloneqq \int\limits_{0}^{T_{\mathrm{ro}}} \frac12 \rho (q_{\mathrm{z},k}(t)) u^2_{\mathrm{\infty}}(q_{\mathrm{z},k}(t))b \lVert \dot{q}_k(t) \rVert  \mathrm{d}t \ ,
\end{equation}
where, by including the dynamic pressure  inside the integral, we account for variability of wind speed and air density along the trajectory.
The parameter $T_{\mathrm{ro}}$ is the reel-out phase duration, and $q_{\mathrm{z},k}$ is the vertical position component of aircraft $k$.

We assume that the force acting on this annulus is the main tether force, and that during the reel-in phase, this force is zero.
The average tether force over one power cycle of period $T$ is then given by:
\begin{equation}
\bar{F}_{\mathrm{t}} \coloneqq \frac{1}{T}\int\limits_0^{T_{\mathrm{ro}}} F_{\mathrm{t}}(t)\mathrm{d}t \ ,
\end{equation}
with the expression of the main tether force $F_{\mathrm{t}}(t)$ in \cite{DeSchutter2019}.

Momentum conservation applied to these average quantities then gives an algebraic equation for the average induction factor $a$, i.e.:
\begin{equation*}
\bar{F}_{\mathrm{t}} = 4 a (1-a)\sum\limits_{k\in\mathcal{K}} A_{\mathrm{s},k} \ .
\end{equation*}
The apparent wind speed that each aircraft $k$ experiences is then given by
\begin{equation}
u_{\mathrm{a},k} \coloneqq (1-a) u_{\infty}(q_{\mathrm{z},k})e_{\mathrm{x}} - \dot{q}
\end{equation}
with $e_{\mathrm{x}}\coloneqq \begin{bmatrix}
1 & 0 & 0 
\end{bmatrix}^{\top}$ the unit vector in the $x$-direction.

Similar to the steady-state models proposed in \cite{DeLellis2018a} and \cite{Kheiri2019}, this induction model assumes that both flight annulus and tether force are perpendicular to the wind speed vector.
This is a crude assumption, given that the main tether elevation angle is intrinsically non-zero.
Nevertheless, this averaged approach gives a first-order account of the time-dependency of the induction for lift-mode systems, which is necessary in particular for lift-mode orbits with short reel-out phases.

By adding the values $A_{\mathrm{s},k}$ in the M-AWES case, we assume that the swept areas of the different aircraft do not overlap.
To avoid double-counting, a no-overlap condition will be enforced as a constraint in the optimal control problem.
For S-AWES, we will check a posteriori that the swept area does not self-overlap.
\section{Problem formulation}\label{sec:OCP}
This section discusses the path constraints used for the simulations in this work, and introduces a constraint which expresses the available flight cylinder as a function of the radius of the corresponding circular ground area.
Then, this section presents a periodic OCP formulation to compute PD-optimal power cycles based on the dynamics presented in the previous section. 
\subsection{Flight envelope}\label{sec:constraints}
Path constraints need to be enforced along the trajectory, to avoid flight envelope violations and to preserve structural integrity of the airframes and the tethers.
More concretely, we impose the following constraints:
\begin{itemize}
	\item Tether stress should not exceed the material yield stress with a safety factor 3.
	\item The tether force should be strictly positive to avoid tether sag and to preserve model validity.
	\item Aircraft roll and pitch angles should be smaller than 90$^{\circ}$ to avoid collision with the tether. Note that for real-world trajectories, a larger safety factor would be appropriate.
	\item The angle-of-attack and side-slip angle of all aircraft are bounded to avoid stall and preserve model validity.
	\item The aileron, elevator and rudder control surfaces and their rates are bounded.
	\item The aircraft should remain above the ground with a safety distance of 100 m.
	\item The tether length is bounded from above by a value $l_{\mathrm{t, max}}$ = 700 m. For the M-AWES, this constraint is active at the end of the reel-out phase.
	\item The induction factor $a$ should be positive and smaller than 0.5 to avoid flow acceleration or reversal.
\end{itemize}
We refer the reader to \cite{DeSchutter2019} for those numerical bound values not mentioned explicitly in this text.
\subsection{Cylindrical flight constraint}
A central feature of vertical M-AWES parks is that each system is assigned an individual, tilted flight cylinder.
The intersection of this flight cylinder with the ground gives a circular area with radius $R$ (= $d/2$).
The flight cylinder has an ellipsoidal cross-section with a major axis length $R$ and a minor axis length $R\sin \theta_{\mathrm{e}}$, with $\theta_{\mathrm{e}}$ the elevation angle of the cylinder.
This elevation angle is included as an optimization variable.
%Increasing $\theta_{\mathrm{e}}$ increases the available flight space, but decreases alignment of the system with the wind, hence decreasing the available wind power.
%
%\begin{figure}
%	\centering
%	\input{figures/side_view_real_AP2_PDGA_radius8.pgf}
%	\input{figures/ellipse_view_real_AP2_PDGA_radius8.pgf}
%\end{figure}

We express the flight cylinder constraint for all aircraft in the system in the following way.
Note that for M-AWES, the constraint is also imposed on the juncture node between main tether and secondary tethers.
First, we rotate the aircraft position into the ellipsoidal cylinder frame:
\begin{align}
\hat{q}_{\mathrm{y},k}& \coloneqq q_{\mathrm{y},k} \\
\hat{q}_{\mathrm{z},k}& \coloneqq  q_{\mathrm{z},k} \cos (\theta_{\mathrm{e}}) -  q_{k,\mathrm{x}} \sin (\theta_{\mathrm{e}}) \ .
\end{align}
Then, we define the constraining ellipse axes as
\begin{align}
 \hat{R}_{\mathrm{y}} &\coloneqq R - b/2 \\
\hat{R}_{\mathrm{z}} &\coloneqq R \sin \theta_{\mathrm{e}} - b/2 \ ,
\end{align}
which ensures that the entire wing with span $b$ remains in the ellipse.
The constraint then reads as
\begin{equation}
	\frac{\hat{q}_{\mathrm{z},k}^2}{\hat{R}_{\mathrm{z}}^2} + 	\frac{\hat{q}_{\mathrm{y},k}^2}{\hat{R}_{\mathrm{y}}^2} \leq 1 \ .
\end{equation}
These flight constraints, together with the constraints mentioned Section \ref{sec:constraints} are summarized by the expression $h(x(t), u(t), z(t), \theta, a) \geq 0$.

In the M-AWES case, we need to ensure that the swept areas of the two connected aircraft do not overlap during the reel-out phase, to preserve model validity.
For the purposes of this work, we therefore propose to pre-structure the M-AWES OCP so that the solution consists of one single loop: half a loop for the reel-out phase, and half a loop for the reel-in phase.
During the reel-out phase, each aircraft is assigned one half of the flight cylinder.
During the reel-in phase, the two aircraft switch flight regions.

Formally, we express the no-overlap constraint for aircraft $k$ as
\begin{align}
h_{\mathrm{no},k}(x, \theta_{\mathrm{e}}, \phi_0) \coloneqq \hat{q}_{\mathrm{z},k}\cos(\phi_0) - \hat{q}_{\mathrm{y},k}\sin (\phi_0)  \ .
\end{align}
The angle $\phi_0$ rotates the intersecting half-plane that divides the flight cylinder in two and can be chosen freely by the optimizer.

For a dual-aircraft system, with aircraft nodes $k\in\mathcal{K} = \{2,3\}$, the no-overlap condition for the two aircraft is combined with a phase-fixing constraint on the tether reel-out speed $\dot{l}_{\mathrm{t}}$:
\begin{equation}\label{eq:no_overlap}
h_{\mathrm{no}}(x, \theta_{\mathrm{e}}, \phi_0) \coloneqq \begin{bmatrix}
h_{\mathrm{no},2}(x, \theta_{\mathrm{e}}, \phi_0)  \\
- h_{\mathrm{no},3}(x, \theta_{\mathrm{e}}, \phi_0) \\
\dot{l}_{\mathrm{t}}
\end{bmatrix} \ .
\end{equation}
This constraint is greater or smaller than zero depending on the phase.The constraints mentioned in this section are summarized by the expression $h(x(t), u(t), z(t), \theta, a) \geq 0$.

\subsection{Optimal control problem}
We can now directly compute periodic flight trajectories that optimize the power density, by solving the following periodic OCP:
\begin{mini*}
	{\substack{x(\cdot),u(\cdot),z(\cdot)\\ \theta, a, T_{\mathrm{ro}}, T_{\mathrm{ri}} \\ R, \theta_{\mathrm{e}}, \phi_0}}{-\frac{1}{T}\int\limits_0^T \rho_{\mathrm{circle}} \frac{P(t)}{\pi R^2} \mathrm{d}t}{\label{eq:POCP}}{}
	\addConstraint{F(\dot{x}(t),x(t),u(t),z(t),\theta)}{=0,}{\ \forall t \in [0, T]}
	\addConstraint{h(x(t),u(t),z(t),\theta, a)}{\geq 0,}{\ \forall t \in [0, T]}
	\addConstraint{h_{\mathrm{no}}(x(t),\theta_{\mathrm{e}}, \phi_0)}{\geq 0,}{\ \forall t \in [0, T_{\mathrm{ro}}]}
	\addConstraint{-h_{\mathrm{no}}(x(t),\theta_{\mathrm{e}}, \phi_0)}{\geq 0,}{\ \forall t \in (T_{\mathrm{ro}}, T]}
	\addConstraint{x(0) - x(T)}{=0}{}
	\addConstraint{\bar{F}_{\mathrm{t}}- 4a(1-a)\sum\limits_{k\in\mathcal{K}} A_{\mathrm{s},k}}{= 0,}{}
\end{mini*}	
where the overall time $T$ is defined as the sum of the reel-out time and the reel-in time, which are free optimization variables: $T \coloneqq T_{\mathrm{ro}} + T_{\mathrm{ri}}$.
The initial and final state of the trajectory are free, but must be equal.
The cost function is chosen so as to maximize the average power output divided by the circular ground area occupied by the system.
This PD is multiplied with $\rho_{\mathrm{circle}}$ to account for packing losses.

The M-AWES OCP is discretized using direct collocation with 40 intervals, and Radau polynomials of degree 4.
For S-AWES, the no-overlap conditions in \eqref{eq:no_overlap} are omitted, but the phase-fixing constraint is retained.
Since the optimal time period is larger for this problem, we increase the number of collocation intervals to 100.

The NLP is formulated in Python using the open-source AWE optimal control framework \texttt{AWEbox} \cite{awebox2020}, which builds on the symbolic framework for algorithmic differentation and nonlinear optimization \texttt{CasADi} \cite{Andersson2019}.
\texttt{AWEbox} solves the NLP with \texttt{IPOPT} \cite{Waechter2006} and the linear solver \texttt{MA57} \cite{HSL}.

\section{Numerical results}\label{sec:results}
This section presents and discusses PD-optimal periodic orbits for M-AWES and S-AWES, both for a small and moderate aircraft size.
In a second step, periodic orbits for each variant are computed for a range of fixed values for the ground circle radius $R$, in order to investigate the trade-off between power density and wing area efficiency.
We conclude with a critical discussion of the obtained results in the light of the modeling assumptions.
\subsection{Optimal power density solutions}
The periodic OCP is solved for both a S-AWES and a M-AWES with two aircraft. 
First we use the small wing span ($b = 5.5$ m) of the original aircraft model.
Then we do the same for an upscaled version of the same model ($b= 26$ m).
Table~\ref{tab:PD_results} summarizes the optimal results for the different variants.
All of the results have a similar optimal elevation angle $\theta_{\mathrm{e}}^{\ast} \approx 40^{\circ}$.
This is close to the theoretically optimal value of $45^{\circ}$ obtained in Section \ref{sec:main_idea}.

\begin{figure}
	\centering
	\input{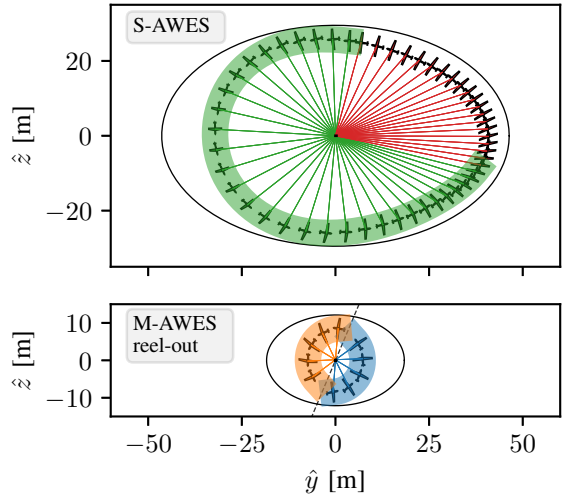}
	\caption{PD-optimal flight trajectories in ellipsoidal coordinates for $b=5.5$ m, with $\theta^{\ast}_{\mathrm{e}} \approx 40^{\circ}$.}\label{fig:small_solution}
\end{figure}

Fig.~\ref{fig:small_solution} shows the optimal trajectories for the small-size S-AWES and M-AWES.
The S-AWES has a rather large optimal ground circle radius $R^{\ast}= 46.5$ m, and very low power output because of the large tether drag losses.
Hence the power density is also impractically low (0.2 MW/km$^2$).
The optimal radius for the M-AWES is a factor 2.5 smaller, at a value only slightly larger than three wing spans.
The dual-aircraft configuration thus allows the system to fly extremely tight circles.
Combined with the efficiency gain due to reduced tether drag and the increased flying altitude, this results in a power density that is 35 times higher (7.3 MW/km$^2$) than for S-AWES.
Optimizing for power density drives this system to make very efficient use of the available airspace: the optimal induction factor $a^{\ast} = 0.21$ is close to the theoretically optimal value of 1/3.
For the S-AWES, induction is almost negligible.

\begin{figure}
	\centering
	\input{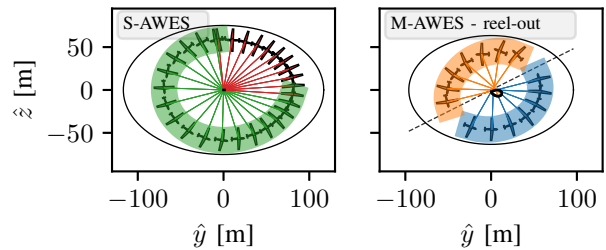}
	\caption{PD-optimal flight trajectories in ellipsoidal coordinates for $b=26.0$ m, with $\theta_{\mathrm{e}}^{\ast} \approx 40^{\circ}$.}\label{fig:moderate_solution}
	\vspace{-0.5cm}
\end{figure}
Fig.~\ref{fig:moderate_solution} shows the optimal trajectories for the moderate-size S-AWES and M-AWES.
The S-AWES solution improves on two fronts compared to the small-size results.
First, since the relative tether drag contribution to the total system drag decreases with increasing aircraft size, the moderate-size system is more efficient and flies at a higher altitude, resulting in a significantly higher power output.
Second, for a more than four times larger aircraft, the ground area radius only increases by a factor of 2.5, the induction factor increases by a factor of 5, and the PD (1.7 MW/km$^2$) increases by a factor larger than 6.

For the M-AWES, the ground radius increases with a factor 5, but the large aircraft size also reduces the impact of the secondary tether drag, which is why the power density increases slightly to 8.4 MW/km$^2$, which is still almost a factor 5 larger than for the moderate-size S-AWES.

Note that while in the envisioned vertical wind farms, M-AWES need an individual flight cylinder to avoid collisions, this need not be the case for S-AWES.
In fact, S-AWES can be packed closer together than done in this work, even with overlapping flight cones when synchronized properly, as proposed in \cite{Faggiani2018}.
From this perspective, the obtained PD results are too pessimistic.

However, the model used in this work neglects wake interaction effects within the farm, which is a good assumption for a vertical M-AWES farm, but not for densely and horizontally packed S-AWES, in particular as they grow larger.
The induction factors obtained in this study are small ($<$ 0.05) but they might still have a non-negligible effect on the power output of downstream systems.
In this sense, the proposed packing densities might not be overly conservative.
%Flying compact trajectories not only reduces the trajectory footprint but also increases efficiency: the shorter the secondary tether lengths, the less dissipation losses in the system.
%This efficiency gain holds approximately up until the point where the harvesting area becomes too small and induction becomes too high.
%The limiting factors are: flight envelope constraints, aircraft control authority over the rotational dynamics.
%In this case, Therefore, if the combination of flight dynamics and constraints would allow for it, the optimal radius would even be lower.
%An alternative way of achieving a more compact flight path would be via a rotary AWE system, where the aircraft wings are linked via a rigid connection \cite{DeSchutter2018}.
%
%
%\begin{figure}
%	\input{figures/ellipse_PDGA_radius0_single.pgf}
%\end{figure}
%%
%\begin{figure}
%	\input{figures/outputs_ellipse_PDGA_radius12.pgf}
%\end{figure}
%
%\begin{figure}
%	\centering
%	\input{figures/ellipse_PDGA_radius0_single_big.pgf}
%\end{figure}

\begin{table}[h]
	\centering
	\caption{PD-optimal (A-D) and practically recommended (E) solution parameters and outputs.}	\label{tab:PD_results}
	%%% \tablesize{} %% You can specify the fontsize here, e.g., \tablesize{\footnotesize}. If commented out \small will be used.
	\begin{tabular}{ccccccc}
		\toprule
		Label & System & $b$ [m] & $a$ [-] & $R$ [m] & PD [$\frac{\text{MW}}{\text{km}^2}$] & $\bar{P}$ [kW]  \\
		\midrule
		A &S-AWES & 5.5 & 0.01 & 46.5 & 0.2  & 2.0\\
		B &M-AWES & 5.5 & 0.21 & 18.4 & 7.3 & 11.0 \\ 
		C &S-AWES & 26.0 & 0.05 & 116.8  & 1.7 & 107.0\\
		D &M-AWES & 26.0  & 0.09 & 96.0  & 8.4 & 347.6 \\
		\midrule
		E & M-AWES & 5.5  & 0.13 & 30.4  & 5.9 & 24.5 \\
		\bottomrule
	\end{tabular}	
\end{table}

\subsection{Trade-off between ground and wing area}
Optimizing for power density results in a suboptimal solution in terms of average power output for a given wing area $\bar{P}_{\mathrm{S}} \coloneqq \bar{P}/(\lvert \mathcal{K} \rvert S)$, with $S$ the aerodynamic surface of a single aircraft in the system.
%When optimizing for the latter, conversion efficiency (in Betz' sense) is typicallly traded for a larger harvesting area, as long as there is a net increase in power output.
In practice, a trade-off between these two objectives needs to be found: for a given power output, we want to both minimize the trajectory footprint and the required wing area.

\begin{figure}
	\input{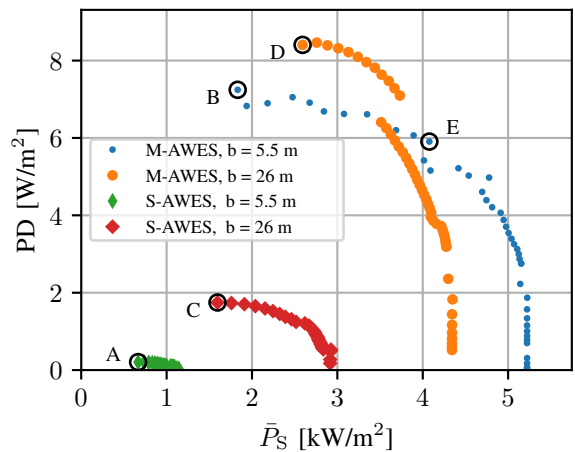}
	\caption{Pareto efficiency front between power density and power per wing area for small- and moderate-size S-AWES and M-AWES.
%		 The labels A and B mark the solutions plotted in Fig.~\ref{fig:small_solution} and the labels C and D  mark those plotted in Fig.~\ref{fig:moderate_solution}.
}\label{fig:pareto_front}
	\vspace{-0.6cm}
\end{figure}

Starting from the PD-optimal solution, the Pareto front between PD and $\bar{P}_{\mathrm{S}}$ is constructed by re-solving the OCP for fixed and increasing values of $R$.
Fig~\ref{fig:pareto_front} shows the result of this parametric sweep for all variants.
After a certain value of $R$, the $\bar{P}_{\mathrm{S}}$-optimal solution is reached and the cylindrical flight constraint becomes inactive for all larger $R$.

For increasing $R$, all system variants are able to increase power output in two ways.
First, the systems fly at lower, more power-optimal elevation angles, down to a value $\theta_{\mathrm{e}} \approx 25^{\circ}$ for all variants.
The larger value of $R$ compensates the minor ellipse axis reduction with $\sin \theta_{\mathrm{e}}$.
Second, the systems fly trajectories with a larger harvesting area, which results in a lower induction factor but overall in a net increase in power.
%When optimizing for average power in an idealized setting, the harvesting area would be infinitely large with zero induction factor \cite{Kheiri2019}.
%taking into account the secondary tether drag losses results in a finite increase in harvesting area.

The M-AWES power output increases by a factor up to 2.9 for the small-size system and by up to a factor of 1.7 for the moderate-size system, at the cost reducing the PD with a factor of 2.6 in both cases.
A good compromise for the small-size system might be PD $\approx$ 6 MW/km$^2$ and $\bar{P}_{\mathrm{S}} \approx$ 4 kW/m$^2$, marked by the label ``E" in Fig.~\ref{fig:pareto_front} and summarized in Table~\ref{tab:PD_results}.
%For the moderate-size S-AWES, a good trade-off might be PD $\approx$ 1.6 MW/km$^2$ and $\bar{P}_{\mathrm{S}} \approx 2.7$.
Interestingly, the small-size M-AWES completely dominates the moderate-size S-AWES by a large margin.
M-AWES based on small aircraft can thus be efficiently deployed both as a single unit for small-scale applications, as well as in AWE farms for utility-scale electricity generation.

%	\begin{itemize}
%	\item Re-solve OCP for increasing values of $R$
%	\item The average power harvesting factor
%	%
%	\begin{equation*}
%	\bar{\zeta} \coloneqq \frac{1}{T}\int\limits_0^T\frac{P(t)}{\sum\limits_{k = 1}^2 q_{\infty, k}(t)S}\mathrm{d}t \ ,
%	\end{equation*}
%	%
%	with $q_{\infty, k}(t)$ the dynamic pressure at the $k$-th kite and $S$ the aerodynamic surface area,
%	is maximized by increasing swept area at the cost of suboptimal $a$.
%\end{itemize}
%A good economic trade-off might be achieved for $\zeta = 18$ and PD $10 \mathrm{W/m}^2$

%\begin{figure}
%	\centering
%	\input{figures/compare_phf_big.pgf}
%\end{figure}
%\begin{figure}
%	\centering
%	\input{figures/ellipse_PDGA_radius12_big.pgf}
%\end{figure}
%\begin{figure}
%	\centering
%	\input{figures/ellipse_PDGA_radius13_single_big.pgf}
%\end{figure}
%The trajectories in this work are characterized by unrealistically high tether accelerations (up to 73 m/s$^2$).
%This is caused by the very short reel-out and reel-in times of the single-loop power cycles.
%We deliberately computed single-loop solutions to be consistent with the proposed induction model.
%Reel-out phases consisting of multiple loops will exhibit a less aggressive reeling behavior, but need more elaborate induction models.

\section{Conclusion}\label{sec:discussion}

In this paper we proposed vertical M-AWES farms with high PD per ground area.
We determined the theoretical potential of these farms and computed and compared detailed PD-optimal flight trajectories for both M-AWES and S-AWES of different sizes.
The achieved PD of the recommended small-size M-AWES design ``E" is significantly lower than the theoretical estimate, by a factor of 7.
A big loss factor is the fact that the optimal flight annulus only covers part of the elliptical cross-section, and thus does not exploit the total available harvesting area.
Future work should explore M-AWES trajectories that use more area of the elliptical cylinders in order to achieve power densities that are closer to what is theoretically possible.
\vspace{-0.2cm}
\section*{Acknowledgements}
This research was supported by DFG via Research Unit FOR 2401 and project 424107692 and by the EU via ELO-X 953348. 
\vspace{-0.5cm}
\bibliographystyle{elsarticle-num}
\bibliography{syscop}

\begin{thebibliography}{10}
\expandafter\ifx\csname url\endcsname\relax
  \def\url#1{\texttt{#1}}\fi
\expandafter\ifx\csname urlprefix\endcsname\relax\def\urlprefix{URL }\fi
\expandafter\ifx\csname href\endcsname\relax
  \def\href#1#2{#2} \def\path#1{#1}\fi

\bibitem{VanZalk2018}
J.~Van~Zalk, P.~Behrens, The spatial extent of renewable and non-renewable
  power generation: A review and meta-analysis of power densities and their
  application in the u.s, Energy Policy 123 (2018) 83--91.

\bibitem{Haas2022}
T.~Haas, J.~De~Schutter, M.~Diehl, J.~Meyers, Large-eddy simulation of airborne
  wind energy farms, Wind Energy Science 7~(3) (2022) 1093--1135.

\bibitem{Zanon2013a}
M.~Zanon, S.~Gros, J.~Andersson, M.~Diehl, Airborne wind energy based on dual
  airfoils, IEEE Transactions on Control Systems Technology 21 (2013)
  1215--1222.

\bibitem{Read2018}
R.~Read, Kite networks for harvesting wind energy, in: Airborne Wind Energy:
  Advances in Technology Development and Research, Springer Singapore, 2018.

\bibitem{DeSchutter2019}
J.~De~Schutter, R.~Leuthold, T.~Bronnenmeyer, R.~Paelinck, M.~Diehl, Optimal
  control of stacked multi-kite systems for utility-scale airborne wind energy,
  in: Proceedings of the IEEE Conference on Decision and Control (CDC), 2019.

\bibitem{Archer2013a}
C.~Archer, An introduction to meteorology for airborne wind energy, in:
  Airborne Wind Energy, Springer Berlin / Heidelberg, 2013.

\bibitem{Malz2019}
E.~C. Malz, J.~Koenemann, S.~Sieberling, S.~Gros, A reference model for
  airborne wind energy systems for optimization and control, Renewable Energy
  140 (2019) 1004--1011.

\bibitem{DeLellis2018a}
M.~{De Lellis}, R.~Reginatto, R.~Saraiva, A.~Trofino, The betz limit applied to
  airborne wind energy, Renewable Energy 127 (2018) 32--40.

\bibitem{Kheiri2019}
M.~Kheiri, V.~S. Nasrabad, F.~Bourgault, A new perspective on the aerodynamic
  performance and power limit of crosswind kite systems, Journal of Wind
  Engineering and Industrial Aerodynamics.

\bibitem{Leuthold2018}
R.~Leuthold, J.~De~Schutter, E.~C. Malz, G.~Licitra, S.~Gros, M.~Diehl,
  Operational regions of a multi-kite awe system, in: European Control
  Conference (ECC), 2018.

\bibitem{awebox2020}
\url{https://github.com/awebox} (2022).

\bibitem{Andersson2019}
J.~Andersson, J.~Gillis, G.~Horn, J.~Rawlings, M.~Diehl, {CasADi} -- a software
  framework for nonlinear optimization and optimal control, Mathematical
  Programming Computation 11~(1) (2019) 1--36.

\bibitem{Waechter2006}
A.~W\"achter, L.~T. Biegler, On the implementation of an interior-point filter
  line-search algorithm for large-scale nonlinear programming, Mathematical
  Programming 106~(1) (2006) 25--57.

\bibitem{HSL}
{HSL}, {A} collection of {F}ortran codes for large scale scientific
  computation. (2011).

\bibitem{Faggiani2018}
P.~Faggiani, R.~Schmehl, Airborne Wind Energy: Advances in Technology
  Development and Research, Springer Singapore, 2018, Ch. Design and Economic
  of a Pumping Kite Wind Park, pp. 391--411.

\end{thebibliography}

\end{document}